\documentclass[a4paper,11pt]{article}
\usepackage{amsmath} 
\usepackage{graphicx, epsfig,float,bbm}
\newcommand{\beq}{\begin{equation}}
\newcommand{\eeq}{\end{equation}}
\newcommand{\beqa}{\begin{eqnarray}}
\newcommand{\eeqa}{\end{eqnarray}}

\newcommand{\half}{{\frac{1}{2}}} 
\newcommand{\ket}[1]{|#1\rangle} 
\newcommand{\bra}[1]{\langle#1|}

\newcommand{\Buzek}{Bu\v{z}ek }

\title{Upper Bounds for the Number of Quantum Clones under Decoherence} 
\author{K. Maruyama and P.L. Knight\\QOLS, The Blackett Laboratory,
Imperial College\\London SW7 2BZ, United Kingdom}
\date{}

\begin{document}
\maketitle
\begin{abstract}
Universal quantum cloning machines (UQCMs), sometimes called quantum cloners,
generate many outputs with identical density matrices, with as close a
resemblance to the input state as is allowed by the basic principles of
quantum mechanics. Any experimental realization of a quantum cloner has
to cope with the effects of decoherence which terminate the coherent evolution
demanded by a UQCM. We examine how many clones can be generated within a
decoherence time.
We compare the time that a quantum cloner implemented with
trapped ions requires to produce $M$ copies from $N$ identical pure state
inputs and the decoherence time during which the probability of spontaneous
emission becomes non-negligible. 
We find a method to construct an $N\rightarrow M$ cloning circuit, and
estimate the number of elementary logic gates required.
It turns out that our circuit is highly vulnerable to spontaneous emission
as the number of gates in the circuit is exponential
with respect to the number of qubits involved.
\end{abstract}

\section{Introduction}
Although the ``no-cloning theorem'' by Wootters and Zurek \cite{wz82} states
that it is impossible to copy arbitrary quantum information perfectly,
quantum mechanics does allow us to make approximate copies.
Since \Buzek and Hillery first presented a unitary transformation,
known as a {\it universal quantum cloning machine} (UQCM),
to make two identical approximate copies of an
input qubit \cite{bh96}, the subject of quantum cloning has been
investigated intensely (for example, see \cite{gisin97, gisin98,
werner98, buzek01}) and experiments have recently been carried out
\cite{huang00, linares02, fasel02}.

One application of quantum cloning is in quantum computation, as copying
is a fundamental process in information processing. It has been shown
that there are
some quantum-computational tasks whose performance can be enhanced
by making use of quantum cloning \cite{galvao00}.
However, physical realizations of quantum circuits
are always fraught with difficulties due to decoherence, as it is
virtually impossible to isolate a quantum system from its environment
perfectly.

Here, we investigate the circuit complexity of universal quantum
cloning and estimate the time $T$ a quantum cloning machine requires to
make $M$ copies of $N$ identical pure state inputs. By comparing $T$
with the decoherence time $\tau_{\mathrm{dec}}$ of the qubits in a
physical system, we can estimate how $M$ or $N$ will be restricted
and how much quantum information can be copied practically. In order to
estimate the decoherence time, one needs to model a specific realization,
and several are possible: cavity QED, parametric down-conversion, and
ion traps. To provide a concrete example, we examine an ion-trap realization
of the UQCM network, and assume all instrumental sources of decoherence
\cite{murao98, james98, schneider98} have been eliminated, leaving us
with spontaneous emission as a well-characterizable source of decoherence.

The analysis of the times $T$ and $\tau_{\mathrm{dec}}$ is analogous to
that reported in \cite{martin9697}, in which upper bounds are determined
for the bit size $L$ of the number to be factorized by using Shor's algorithm.

In the next section, we briefly review the ideas of quantum cloning,
and then, in Section \ref{sec_circuit}, count the number of elementary
logical operations needed to build a quantum circuit for $N\rightarrow M$
cloning. In Section \ref{dec_time}, we compare the cloning time $T$
with the decoherence time $\tau_{\mathrm{dec}}$ of the quantum circuit
when it is implemented with trapped ions, as in \cite{martin9697}. Other
sources of decoherence (vibrational quanta damping, heating, inhomogeneous
trap fields, and laser fluctuations) will also lead to limitations in the
number of good quality copies such a UQCM can generate.

\section{Quantum Cloning Transformation}\label{sec_trans}
The task of a general universal quantum cloning machine is to copy
$N$ identical pure states, described as
$\rho^{\mathrm{in}}=(\ket{\psi}\bra{\psi})^{\otimes N}$, into
$M$-particle output states ($M>N$), $\rho^{\mathrm{out}}$, with the
following conditions:

\begin{enumerate}
 \item The reduced density operators for any one of the $M$ outputs are
identical to each other, i.e.,
 \begin{equation}
  \rho^{\mathrm{out}}_i=\rho^{\mathrm{out}}_j,
 \end{equation}
 where $\rho^{\mathrm{out}}_i$ is a reduced density operator with respect
 to the $i$-th particle.

 \item The quality of the copies does not depend on the input states,
i.e. the {\it fidelity} between the input and the output states is
independent of the input state. The fidelity $F$ is defined by
$F=\bra{\psi}\rho^{\mathrm{out}}_i\ket{\psi}$. The word ``universal''
refers to this condition.

 \item The copies are as close as possible to the input state as a natural
requirement for a cloning machine. Thus, the fidelity $F$ should be as close
as possible to 1.
\end{enumerate}

Here, we consider only a two-dimensional system (qubit) as a
physical system, as systems of higher dimension are beyond the scope of
current candidates for the realizations of quantum circuits.

The cloning transformation of \Buzek and Hillery \cite{bh96} makes two
copies from one original qubit. It is written as
\begin{eqnarray} \label{ucm01}
  \ket{0}_{a}\ket{0}_{b}\ket{0}_{x}& \rightarrow &
   \sqrt{\frac{2}{3}}\ket{0}_{a}\ket{0}_{b}\ket{0}_{x}
   +\sqrt{\frac{1}{6}}(\ket{0}_{a}\ket{1}_{b}
   +\ket{1}_{a}\ket{0}_{b})\ket{1}_{x} \nonumber \\
  \ket{1}_{a}\ket{0}_{b}\ket{0}_{x}& \rightarrow &
   \sqrt{\frac{2}{3}}\ket{1}_{a}\ket{1}_{b}\ket{1}_{x}
   +\sqrt{\frac{1}{6}}(\ket{0}_{a}\ket{1}_{b}
   +\ket{1}_{a}\ket{0}_{b})\ket{0}_{x}.
\end{eqnarray}
In Eq. (\ref{ucm01}), the first qubit with subscript $a$ is the state
to be copied, the second one labelled $b$ is the {\it blank paper} that becomes
one of the copies after the process, and the qubit with $x$ is an
ancilla bit which can be regarded as the state of the machine.
The fidelity of this process is found to be $\frac{5}{6}$, which is
independent of the input state, as desired.

\begin{figure}
 \begin{center}
  \includegraphics[scale=0.475]{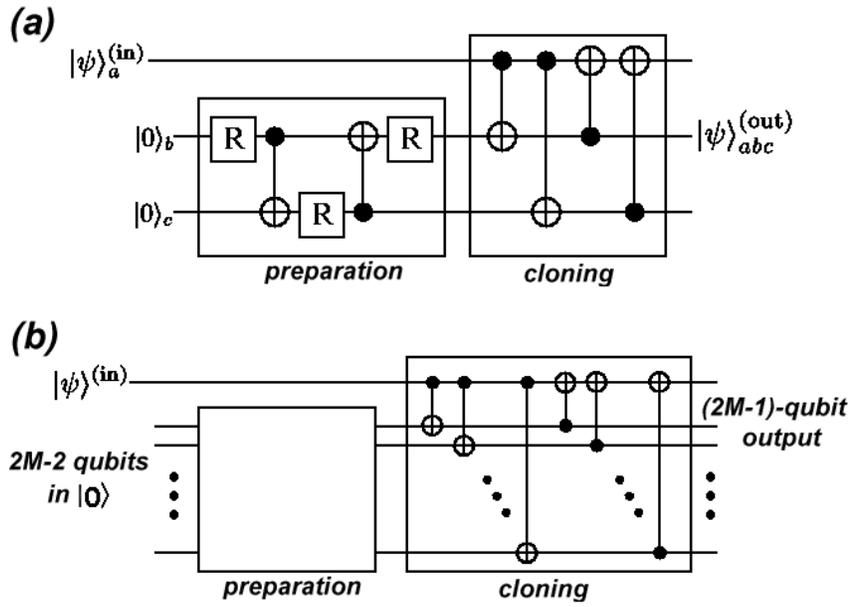}
  \caption{Quantum circuits for (a) $1\rightarrow 2$ UQCM, and
(b) $1\rightarrow M$ UQCM.
These circuits consist of two stages, namely the preparation
and the copying stages. The preparation stage provides appropriate
amplitudes for orthonormal bases according to Eqs. (2) or (4).
The cloning stage permutes those amplitudes among all qubits, entangling
them([13, 14]).}\label{uqcmnet}
 \end{center}
\end{figure}

\Buzek et al. also presented a way to construct a quantum network for this UQCM
\cite{buzek97} (Figure \ref{uqcmnet}(a)). In Figure \ref{uqcmnet}(a), R is a
single qubit gate which rotates the basis vectors by an angle $\theta$ as
\begin{eqnarray} \label{rotation01}
 R(\theta)\ket{0}&=& \cos\theta\ket{0}+\sin\theta\ket{1}, \nonumber \\
 R(\theta)\ket{1}&=&-\sin\theta\ket{0}+\cos\theta\ket{1},
\end{eqnarray}
and $\bullet$ and $\oplus$ symbols connected with a
line denote a controlled NOT gate (CNOT) with $\bullet$ and $\oplus$ as
a control bit and a target bit, respectively. By adjusting the rotation
angles of three single qubit gates, any two of the three states at the
output represent copies of the qubit $a$.

Eq. (\ref{ucm01}) was generalized by Gisin and Massar \cite{gisin97}
to produce $M$ copies out of $N$ inputs. This transformation is
described by
\begin{eqnarray} \label{guqcm}
  U_{N,M}\ket{N\psi}\ket{0}^{\otimes 2(M-N)} &=&
  \sum_{j=0}^{M-N}\alpha_j\ket{(M-j)\psi,j\psi^{\perp}}\otimes
  R_j(\psi), \\
  \alpha_j &=& \sqrt{\frac{N+1}{M+1}}\sqrt{\frac{(M-N)!(M-j)!}
   {(M-N-j)!M!}}, \nonumber \\
  R_j(\psi) &=& \ket{(M-N-j)\psi^*, j(\psi^*)^\perp}, \nonumber
\end{eqnarray}
where $\ket{N\psi}$ is the input state consisting of $N$ qubits, all in
the state $\ket{\psi}$. $\ket{(M-j)\psi,j\psi^\perp}$ is the symmetric
and normalized state with $M-j$ qubits in the state $\ket{\psi}$ and $j$
qubits in the orthogonal state $\ket{\psi^\perp}$. $R_j(\psi)$ represent
the internal state of the machine and $R_j(\psi)\perp
R_k(\psi)$ holds for all $j\ne k$.

\section{Generic Quantum Cloning Circuit}\label{sec_circuit}
\begin{figure}
 \begin{center}
  \includegraphics[scale=0.44]{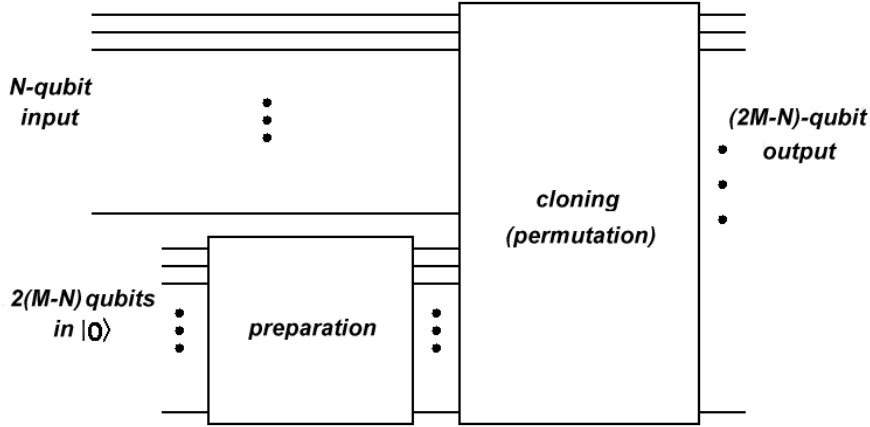}
  \caption{Quantum circuit for $N\rightarrow M$ UQCM? It turns out that
this cloning circuit does work properly for any $M$ when $N=1$, $N=2$,
and many other $N$ and $M$ as long as the condition (\ref{cond})
is fulfilled. If we are allowed to introduce additional auxiliary qubits
inside the cloner, this circuit works for all combinations of $N$ and
$M (>N)$.}
 \label{n2mnet}
 \end{center}
\end{figure}

\Buzek et al. constructed a quantum circuit for $1\rightarrow M$ UQCM
which explicitly realizes Eq. (\ref{guqcm}) by generalizing the
corresponding circuit for the $1\rightarrow 2$ case \cite{buzek98}.
Figure \ref{uqcmnet}(b) shows their circuit for $1\rightarrow M$
cloning machine. It is basically a natural extension of the
$1\rightarrow 2$ circuit, consisting a preparation stage and a cloning
(permutation) stage. Let us consider if it is possible to extend this
$1\rightarrow M$ cloning circuit further to an $N\rightarrow M$ cloning
circuit, by making up a circuit in Figure \ref{n2mnet} in which the
preparation stage is built up with both single and two qubit operations
and the cloning stage is a sequence of CNOT or similar multi-qubit
operation gates.

Eq. (\ref{guqcm}) gives us a few hints on the possible construction
of the quantum circuit for an $N\rightarrow M$ UQCM.
First, $M-1$ qubits are necessary to represent the internal states
of the machine. The total number of qubits needed to implement
this transformation is $2M-N$.

Second, every basis of the form of
$\ket{(M-j)\psi,j\psi^{\perp}}$ or $R_j(\psi)$ in Eq. (\ref{guqcm})
is a basis in the symmetrical subspace of the Hilbert space
$\mathcal{H}^{\otimes M}$ or $\mathcal{H}^{\otimes (M-N)}$, where
$\mathcal{H}$ is the space spanned by $\ket{0}$ and $\ket{1}$.
Although the symmetrical subspace is a rather small subspace in
the whole Hilbert space, almost all computational bases are involved
in Eq. (\ref{guqcm}), especially when $N$ is small.

Therefore, almost $2^{2M-N}$ amplitudes need to be distributed to
appropriate computational bases to perform the transformation (\ref{guqcm}),
although the number of distinct amplitudes, $\alpha_j$, is only $M-N+1$.
The fact that such a great number of amplitudes are necessary imposes
a certain condition upon the values of $N$ and $M$ in order for an
$N\rightarrow M$ cloning machine in Figure \ref{n2mnet} to work properly.

Because the task of the cloning stage which consists of a sequence
of CNOTs is to permute the amplitudes
of all basis vectors which span the whole Hilbert space, all the
amplitudes in Eq. (\ref{guqcm}), $\alpha_j$, must be provided by the
preparation stage. Therefore, the preparation stage should be able to
generate an entangled $2(M-N)$-qubit state with a set of any real
amplitudes that may appear in Eq. (\ref{guqcm}) by adjusting the
rotation angles of each single qubit gate.

For example, in the $1\rightarrow 2$
UQCM (Eq. (\ref{ucm01}) and Figure \ref{uqcmnet}(a)), the amplitudes appearing
in the output states are $\sqrt{\frac{2}{3}}$, $\sqrt{\frac{1}{6}}$, and
$\sqrt{\frac{1}{6}}$. Thus, the preparation stage needs to provide
these three coefficients and it turns out that, with the configuration
of the cloning stage of Figure \ref{uqcmnet}(a), the two-qubit state emerging
from the preparation stage should be
\begin{equation} \label{12prep}
 \ket{\psi}^{\mathrm{prep}}=\sqrt{\frac{2}{3}}\ket{00}+\sqrt{\frac{1}{6}}
 \ket{01}+\sqrt{\frac{1}{6}}\ket{11},
\end{equation}
when the third qubit is used as the internal state of the machine.
The rotation angles of the single qubit gates in this stage are
determined accordingly.

However, this scheme only works when the number of
distinct computational bases in Eq. (\ref{guqcm}) is smaller than the
dimension of the Hilbert space that the preparation stage deals with.
Otherwise, the preparation stage cannot generate enough amplitudes required
by Eq. (\ref{guqcm}). This condition is written as
\begin{equation}\label{cond}
 \sum_{k=0}^{M-N}{M \choose k}{M-N \choose k} \le 2^{2(M-N)},
\end{equation}
where the {\small LHS} represents the number of bases that appear
in Eq. (\ref{guqcm}) and the {\small RHS} is the dimension of the
Hilbert space where qubits in the preparation stage lie.

Eq. (\ref{cond}) is fulfilled for any $M$ when $N=1$, which is
the case of Figure \ref{uqcmnet}(b), and when $N=2$. Still, it becomes
rather complicated when it comes to other combinations of $N$ and $M$.
An example in which we can see the violation of Eq. (\ref{cond}) clearly
is the case of $2N=M$, where the {\small LHS} of Eq. (\ref{cond}) for
$N\gg 1$ can be estimated as
\begin{equation}
 \sum_{k=0}^{N}{2N \choose k}{N \choose k}={3N \choose N} \simeq
 \sqrt{\frac{3}{4\pi}}\left(\frac{27}{4}\right)^N \frac{1}{\sqrt N}.
\end{equation}
This always exceeds the corresponding {\small RHS} of Eq. (\ref{cond}),
$2^{2N}$. Nevertheless, Eq. (\ref{cond}) is satisfied for $M$ which
are sufficiently large compared with $N$, since its {\small LHS}
is approximately $2^{2M-1}/\sqrt{\pi M}$,
which is smaller than the {\small RHS} when $M>\pi^{-1}2^{4N-2}$.

Note that the condition (\ref{cond}) is not necessary if we are allowed to
make use of more auxiliary qubits inside a cloning machine. Providing
$2M-N$ qubits at most, instead of $2(M-N)$, enables the preparation stage to
generate enough number of amplitudes for the cloning transformation.
Then the cloning stage can complete the whole process by allocating
those amplitudes to appropriate bases, leaving the auxiliary qubits
disentangled, in state $\ket{0}$, from the legitimate $2M-N$ output qubits.
Introducing these auxiliary qubits does not affect the estimation of
the number of gates, which is discussed in the following.

Let us now count the number of elementary gates in the quantum cloning
circuit, especially CNOT gates, as a CNOT gate takes
a longer time to be performed than a single qubit gate \cite{martin9697}.

The preparation stage should be designed so that it generates an arbitrary
set of real amplitudes whose number is written by the {\small LHS} of Eq. (\ref{cond}).
As far as the quantum cloning transformation Eq. (\ref{guqcm}) is concerned,
the amplitudes should not necessarily be arbitrary as they are specifically
described as in Eq. (\ref{guqcm}). However, in order to keep the
full ``controllability'' on the output states, as in the case of $1\rightarrow 2$
UQCM, we assume that the preparation stage can generate arbitrary superpositions
of $2^{d_{\mathrm{prep}}}$ bases with real amplitudes, where
$d_{\mathrm{prep}}=2^{2(M-N)}$ is the dimension of the $2(M-N)$-qubit
Hilbert space.

The transformation that the preparation stage performs can be written as
$\ket{00\cdots 0}\rightarrow \sum_{k=0}^{d_{\mathrm{prep}}-1} c_k\ket{k}$, where
$c_k$ are real numbers complying with the normalization
condition $\sum_k c_k^2=1$. This is a unitary transformation
which can be described by a $d_{\mathrm{prep}}\times d_{\mathrm{prep}}$
matrix, on the $d_{\mathrm{prep}}$-dimensional initial state vector
$(1, 0, \cdots, 0)^T$, and its components are given by $u_{m1}=c_m$ and the
rest of them are arbitrary for our purpose.

It is known that an outright circuit implementation of a $d\times d$ unitary
matrix requires, in general, $O(d^2(\log d)^2)$ elementary operations \cite{nielsen00}.
However, a more efficient circuit to create an arbitrary quantum superposition
starting from $\ket{00\cdots 0}$ has been proposed in \cite{long01} and
its complexity is given by $O(d(\log d)^2)$.  Thus we simply
take $O(d_{\mathrm{prep}}(\log d_{\mathrm{prep}})^2)$ as the
number of CNOTs in the preparation stage of a UQCM in the following
calculations.

The cloning stage is also generically hard to construct. The only exception
we know of is the case of $1\rightarrow M$ UQCM, which can be realized
by the circuit in Figure \ref{uqcmnet}(b).
In a more generic $N\rightarrow M$ case,
we can build up a circuit as follows.
As mentioned above, the cloning stage only permutes all bases with non-zero
amplitudes among all computational bases of $\mathcal{H}^{\otimes (2M-N)}$.
Thus, we can estimate the number of gates with only the knowledge of
number of bases involved, even if we have no information on the
actual permutation, i.e. which basis goes to which. 

\begin{table}
\caption{The permutation of computational bases performed in the
cloning stage of the $1\rightarrow 2$ quantum cloning circuit. The choice of
bases on the left hand side follows Eq. (5).}\label{12perm}
\begin{center}
\begin{tabular}{ccc}\hline
 Initial basis & $\rightarrow$ & Final basis \\ \hline 
 $\ket{000}$ & $\rightarrow$ & $\ket{000}$ \\
 $\ket{001}$ & $\rightarrow$ & $\ket{101}$ \\
 $\ket{011}$ & $\rightarrow$ & $\ket{011}$ \\
 $\ket{100}$ & $\rightarrow$ & $\ket{111}$ \\
 $\ket{101}$ & $\rightarrow$ & $\ket{010}$ \\
 $\ket{111}$ & $\rightarrow$ & $\ket{100}$ \\ \hline
\end{tabular}
\end{center}
\end{table}

Let us take the $1\rightarrow 2$ quantum cloning circuit as an example
to simplify our description of its construction, although its efficient
circuit is already given in Figure \ref{uqcmnet}. Table \ref{12perm}
shows all the necessary permutations of bases to complete the transformation
from Eq. (\ref{12prep}) to Eq. (\ref{ucm01}). This permutation can be
implemented as a circuit by carrying out each transformation one by one.
Since $\ket{010}$ is not used in the preparation state, $\ket{101}\rightarrow
\ket{010}$ should be performed first, otherwise any other destination state
may be used as a preparation state later and thus the preceding transformation
comes to naught. A basis which is exempted from being used in the preparation,
such as $\ket{010}$ in this case, always exists in a more general
$N\rightarrow M$ case as the equality in condition (\ref{cond}) never holds
for $N, M\in \mathrm{I\!\!\!\;N}$. Hence, the appropriate order of transformation is
$\ket{101}\rightarrow\ket{010}, \ket{001}\rightarrow\ket{101},
\ket{111}\rightarrow\ket{001}, \ket{100}\rightarrow\ket{111},
\ket{001}\rightarrow\ket{100}$, where $\ket{001}$ is used as a buffer
in the last three operations, as it is not used in the final state,
to swap $\ket{100}$ and $\ket{111}$.

\begin{figure} 
 \begin{center}
  \includegraphics[scale=0.48]{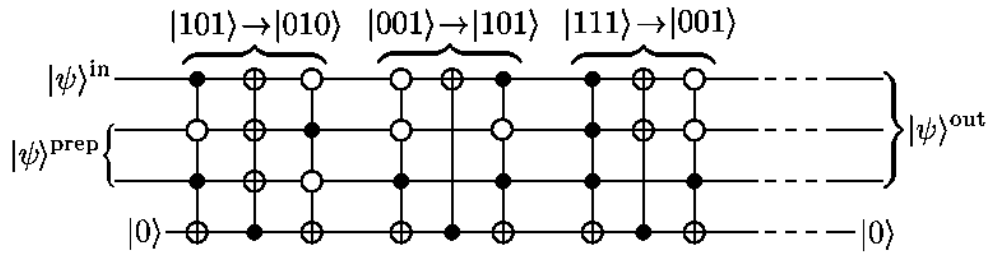}
  \caption{The quantum circuit for the permutation of bases. This figure
depicts a circuit for the cloning stage of $1\rightarrow 2$ UQCM.
An additional ancilla qubit is introduced as a {\it flag} to flip bits.
Although there exists a much more efficient circuit
for the same process, which is shown in Figure \ref{uqcmnet}(a), this
circuit can be applied to the UQCM with more input qubits, as in Figure
\ref{2to4permnet}.}\label{12net} 
 \end{center}
\end{figure}

Figure \ref{12net} shows a general way to permute bases. In this
figure, an unfilled circle, $\circ$, denotes a control bit in-between
two NOT gates. Unlike a filled circle $\bullet$, $\circ$ activates
connected operations when the bit value at $\circ$ is $0$.

Here, an ancilla qubit is introduced in order to flip the incoming
binary numbers according to the permutation required for a UQCM, as shown
in Table \ref{12perm}. Finding a circuit for the basis permutation
without ancillas is generally very hard, hence we make use of it here for
convenience. The ancilla
is prepared by the UQCM in a state $\ket{0}$ and it ends up in the same
state $\ket{0}$ after the whole process. As the ancilla qubit is
necessarily disentangled by the end of the process, we do not have to
take it as a part of the output state.

In the earlier paragraph of this section, we have identified the
generic features required of a copying circuit. For concreteness,
it is important to show a specific example and in the following we
do this for a $2\rightarrow 4$ cloner. Our circuit is not unique
and other examples of course can be found.

The transformation of the state $\ket{00}$ by the $2\rightarrow 4$ quantum
cloning can be written from Eq. (\ref{guqcm}) as
\begin{eqnarray}\label{2to4uqcm}
  \ket{00}\ket{0000} &\rightarrow&
  \sqrt{\frac{3}{5}}\ket{0000}\ket{11} \nonumber \\
  & & +\sqrt{\frac{3}{80}}(\ket{0001}+\ket{0010}+\ket{0100}+\ket{1000})
  (\ket{01}+\ket{10}), \nonumber \\
  & & +\sqrt{\frac{1}{60}}(\overbrace{\ket{0011}+\cdots
  +\ket{1100}}^{6 \:\mathrm{bases}})\ket{00},
\end{eqnarray}
and the transformation for $\ket{11}\ket{0000}$ can be obtained by flipping
every bit in Eq. (\ref{2to4uqcm}).

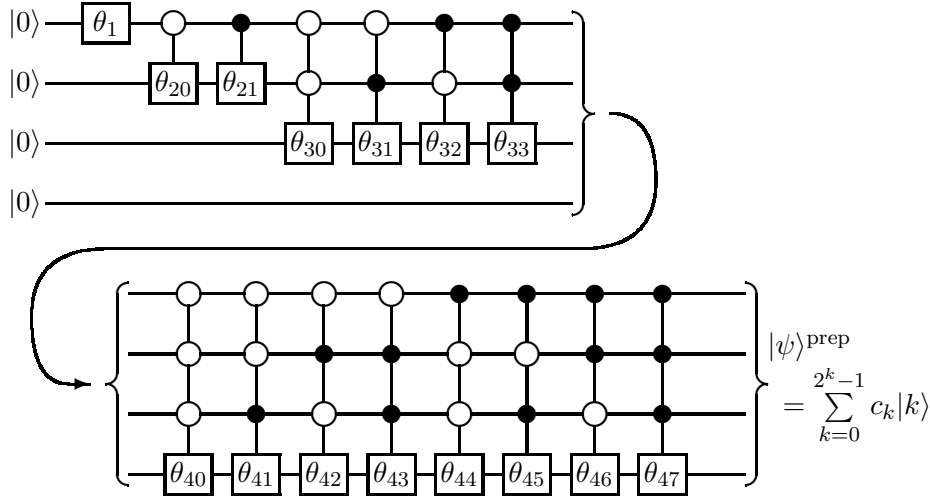
\begin{figure}
\begin{picture}(150,70)(1.5,0)
\thicklines
\put(5,63){$\ket{0}$}
\put(10,64){\line(1,0){5}}
\put(15,61.5){\framebox(6,5){$\theta_1$}}
\put(21,64){\line(1,0){4.5}}
\put(27,64){\circle{3}}
\put(28.5,64){\line(1,0){15}}
\put(36,64){\circle*{2.5}}
\put(45,64){\circle{3}}
\put(46.5,64){\line(1,0){6}}
\put(54,64){\circle{3}}
\put(55.5,64){\line(1,0){24.5}}
\put(63,64){\circle*{2.5}}
\put(72,64){\circle*{2.5}}

\put(27,58.5){\line(0,1){4}}
\put(36,58.5){\line(0,1){5.5}}
\put(45,57.5){\line(0,1){5}}
\put(54,50.5){\line(0,1){12}}
\put(63,57.5){\line(0,1){6.5}}
\put(72,50.5){\line(0,1){13.5}}

\put(5,55){$\ket{0}$}
\put(10,56){\line(1,0){14}}
\put(24,53.5){\framebox(6,5){$\theta_{20}$}}
\put(30,56){\line(1,0){3}}
\put(33,53.5){\framebox(6,5){$\theta_{21}$}}
\put(39,56){\line(1,0){4.5}}
\put(45,56){\circle{3}}
\put(46.5,56){\line(1,0){15}}
\put(54,56){\circle*{2.5}}
\put(63,56){\circle{3}}
\put(64.5,56){\line(1,0){15.5}}
\put(72,56){\circle*{2.5}}

\put(5,47){$\ket{0}$}
\put(10,48){\line(1,0){32}}
\put(42,45.5){\framebox(6,5){$\theta_{30}$}}
\put(48,48){\line(1,0){3}}
\put(51,45.5){\framebox(6,5){$\theta_{31}$}}
\put(57,48){\line(1,0){3}}
\put(60,45.5){\framebox(6,5){$\theta_{32}$}}
\put(66,48){\line(1,0){3}}
\put(69,45.5){\framebox(6,5){$\theta_{33}$}}
\put(75,48){\line(1,0){5}}
\put(45,50.5){\line(0,1){4}}
\put(63,50.5){\line(0,1){4}}
\put(5,39){$\ket{0}$}
\put(10,40){\line(1,0){70}}

\put(80,40){\oval(3,3)[br]}
\put(81.5,40){\line(0,1){10.5}}
\put(81.5,53.5){\line(0,1){10.5}}
\put(80,64){\oval(3,3)[tr]}
\put(83,50.5){\oval(3,3)[tl]}
\put(83,53.5){\oval(3,3)[bl]}

\put(21,28){\line(1,0){6.5}}
\put(29,28){\circle{3}}
\put(30.5,28){\line(1,0){6}}
\put(38,28){\circle{3}}
\put(39.5,28){\line(1,0){6}}
\put(47,28){\circle{3}}
\put(48.5,28){\line(1,0){6}}
\put(56,28){\circle{3}}
\put(57.5,28){\line(1,0){45.5}}
\put(65,28){\circle*{2.5}}
\put(74,28){\circle*{2.5}}
\put(83,28){\circle*{2.5}}
\put(92,28){\circle*{2.5}}
\put(29,21.5){\line(0,1){5}}
\put(29,13.5){\line(0,1){5}}
\put(29,6.5){\line(0,1){4}}
\put(38,21.5){\line(0,1){5}}
\put(38,6.5){\line(0,1){12}}
\put(47,13.5){\line(0,1){13}}
\put(47,6.5){\line(0,1){4}}
\put(56,6.5){\line(0,1){20}}
\put(65,21.5){\line(0,1){6.5}}
\put(65,13.5){\line(0,1){5}}
\put(65,6.5){\line(0,1){4}}
\put(74,21.5){\line(0,1){6.5}}
\put(74,6.5){\line(0,1){12}}
\put(83,13.5){\line(0,1){14.5}}
\put(83,6.5){\line(0,1){4}}
\put(92,6.5){\line(0,1){21.5}}

\put(21,20){\line(1,0){6.5}}
\put(29,20){\circle{3}}
\put(30.5,20){\line(1,0){6}}
\put(38,20){\circle{3}}
\put(39.5,20){\line(1,0){24}}
\put(47,20){\circle*{2.5}}
\put(56,20){\circle*{2.5}}
\put(65,20){\circle{3}}
\put(66.5,20){\line(1,0){6}}
\put(74,20){\circle{3}}
\put(75.5,20){\line(1,0){27.5}}
\put(83,20){\circle*{2.5}}
\put(92,20){\circle*{2.5}}

\put(21,12){\line(1,0){6.5}}
\put(29,12){\circle{3}}
\put(30.5,12){\line(1,0){15}}
\put(38,12){\circle*{2.5}}
\put(47,12){\circle{3}}
\put(48.5,12){\line(1,0){15}}
\put(56,12){\circle*{2.5}}
\put(65,12){\circle{3}}
\put(66.5,12){\line(1,0){15}}
\put(74,12){\circle*{2.5}}
\put(83,12){\circle{3}}
\put(84.5,12){\line(1,0){18.5}}
\put(92,12){\circle*{2.5}}

\put(21,4){\line(1,0){5}}
\put(26,1.5){\framebox(6,5){$\theta_{40}$}}
\put(32,4){\line(1,0){3}}
\put(35,1.5){\framebox(6,5){$\theta_{41}$}}
\put(41,4){\line(1,0){3}}
\put(44,1.5){\framebox(6,5){$\theta_{42}$}}
\put(50,4){\line(1,0){3}}
\put(53,1.5){\framebox(6,5){$\theta_{43}$}}
\put(59,4){\line(1,0){3}}
\put(62,1.5){\framebox(6,5){$\theta_{44}$}}
\put(68,4){\line(1,0){3}}
\put(71,1.5){\framebox(6,5){$\theta_{45}$}}
\put(77,4){\line(1,0){3}}
\put(80,1.5){\framebox(6,5){$\theta_{46}$}}
\put(86,4){\line(1,0){3}}
\put(89,1.5){\framebox(6,5){$\theta_{47}$}}
\put(95,4){\line(1,0){8}}

\put(21,4){\oval(3,3)[bl]}
\put(19.5,4){\line(0,1){10.5}}
\put(19.5,17.5){\line(0,1){10.5}}
\put(21,28){\oval(3,3)[tl]}
\put(18,14.5){\oval(3,3)[tr]}
\put(18,17.5){\oval(3,3)[br]}

\qbezier(85,52)(91,52)(91,43)
\qbezier(81,34)(91,34)(91,43)
\put(18,34){\line(1,0){63}}
\put(13,16){\vector(1,0){3}}
\qbezier(13,16)(8,16)(8,25)
\qbezier(8,25)(8,34)(18,34)

\put(106,20){$\ket{\psi}^{\mathrm{prep}}$}
\put(108,12){$=\sum\limits_{k=0}^{2^k-1}c_k\ket{k}$}

\put(103,4){\oval(3,3)[br]}
\put(104.5,4){\line(0,1){10.5}}
\put(104.5,17.5){\line(0,1){10.5}}
\put(103,28){\oval(3,3)[tr]}
\put(106,14.5){\oval(3,3)[tl]}
\put(106,17.5){\oval(3,3)[bl]}
\end{picture}
\caption{The preparation stage for $2\rightarrow 4$ quantum cloning
circuit. Angles $\theta$ in boxes are given by Eqs. (\ref{thetas}).}\label{2to4prepnet}
\end{figure}

As mentioned earlier in this section, we need four qubits for the
preparation stage in addition to two input qubits, thus, six qubits in total.
The preparation stage generates all amplitudes for the 15 bases appearing in
Eq. (\ref{2to4uqcm}). The output from the preparation stage is, for example,
\begin{eqnarray}\label{2to4preptrans}
  \ket{0000} \rightarrow \ket{\psi}^{\mathrm{prep}} &=&
  \sqrt{\frac{3}{5}}\ket{0000}+\sqrt{\frac{3}{80}}(\overbrace{\ket{0001}+\cdots +\ket{1000}}
  ^{8 \:\mathrm{bases}}) \nonumber \\
  & & +\sqrt{\frac{1}{60}}(\overbrace{\ket{1001}+\cdots +\ket{1110}}
  ^{6 \:\mathrm{bases}}).
\end{eqnarray}
The quantum circuit to carry out this transformation is given in
\cite{long01} and is also depicted in Figure \ref{2to4prepnet}.
The boxes represent a single qubit operation $U_{\theta}$ of the form
\begin{equation}\label{rotation02}
 U_{\theta}=\left(
 \begin{array}{cc}
   \cos\theta & \sin\theta \\ \sin\theta & -\cos\theta
 \end{array}\right),
\end{equation}
and all $\theta$ in Figure \ref{2to4prepnet} are given by
\begin{equation}
  \theta_1=\tan^{-1} \sqrt{\frac{11}{69}}, \:\:\:
  \theta_{20}=\tan^{-1}{\sqrt{\frac{4}{19}}}, \:\:\:
  \theta_{21}=\tan^{-1}{\sqrt{\frac{4}{7}}},\nonumber
\end{equation}
\begin{equation}
  \theta_{30}=\tan^{-1}{\sqrt{\frac{2}{17}}}, \:\:
  \theta_{31}=\frac{\pi}{4}, \:\: \theta_{32}=\tan^{-1}{\sqrt{\frac{8}{13}}}, \:\:
  \theta_{33}=\tan^{-1}{\frac{1}{\sqrt{2}}},\nonumber
\end{equation}
\begin{equation}\label{thetas}
  \theta_{40}=\tan^{-1}{\frac{1}{4}},
  \theta_{41}=\theta_{42}=\theta_{43}=\theta_{45}=\theta_{46}=\frac{\pi}{4},
  \theta_{44}=\tan^{-1}{\frac{2}{3}}, \theta_{47}=0,
\end{equation}

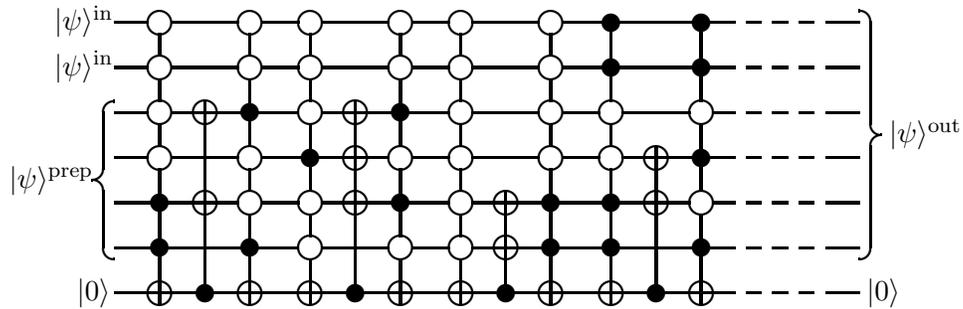
\begin{figure}
\begin{picture}(150,44)(-5,0)
\thicklines
\put(2,40){$\ket{\psi}^{\mathrm{in}}$}
\put(2,34){$\ket{\psi}^{\mathrm{in}}$}
\put(-4,19){$\ket{\psi}^{\mathrm{prep}}$}
\put(5,4){$\ket{0}$}
\put(110,4){$\ket{0}$}
\put(113,25){$\ket{\psi}^{\mathrm{out}}$}
\put(16,5){\circle{3}}
\put(22,5){\circle*{2.5}}
\put(28,5){\circle{3}}
\put(36,5){\circle{3}}
\put(42,5){\circle*{2.5}}
\put(48,5){\circle{3}}
\put(56,5){\circle{3}}
\put(62,5){\circle*{2.5}}
\put(68,5){\circle{3}}
\put(76,5){\circle{3}}
\put(82,5){\circle*{2.5}}
\put(88,5){\circle{3}}

\put(10,41){\line(1,0){4.5}}
\put(16,41){\circle{3}}
\put(17.5,41){\line(1,0){9}}
\put(28,41){\circle{3}}
\put(29.5,41){\line(1,0){5}}
\put(36,41){\circle{3}}
\put(37.5,41){\line(1,0){9}}
\put(48,41){\circle{3}}
\put(49.5,41){\line(1,0){5}}
\put(56,41){\circle{3}}
\put(57.5,41){\line(1,0){9}}
\put(68,41){\circle{3}}
\put(69.5,41){\line(1,0){23}}
\put(76,41){\circle*{2.5}}
\put(88,41){\circle*{2.5}}

\put(22,29){\circle{3}}
\put(22,17){\circle{3}}
\put(42,29){\circle{3}}
\put(42,23){\circle{3}}
\put(42,17){\circle{3}}
\put(62,17){\circle{3}}
\put(62,11){\circle{3}}
\put(82,23){\circle{3}}
\put(82,17){\circle{3}}

\put(16,3.5){\line(0,1){18}}
\put(16,24.5){\line(0,1){3}}
\put(16,30.5){\line(0,1){3}}
\put(16,36.5){\line(0,1){3}}
\put(22,5){\line(0,1){25.5}}
\put(28,3.5){\line(0,1){12}}
\put(28,18.5){\line(0,1){3}}
\put(28,24.5){\line(0,1){9}}
\put(28,36.5){\line(0,1){3}}
\put(36,3.5){\line(0,1){6}}
\put(36,12.5){\line(0,1){3}}
\put(36,18.5){\line(0,1){9}}
\put(36,30.5){\line(0,1){3}}
\put(36,36.5){\line(0,1){3}}
\put(42,5){\line(0,1){25.5}}
\put(48,3.5){\line(0,1){6}}
\put(48,12.5){\line(0,1){9}}
\put(48,24.5){\line(0,1){9}}
\put(48,36.5){\line(0,1){3}}
\put(56,3.5){\line(0,1){6}}
\put(56,12.5){\line(0,1){3}}
\put(56,18.5){\line(0,1){3}}
\put(56,24.5){\line(0,1){3}}
\put(56,30.5){\line(0,1){3}}
\put(56,36.5){\line(0,1){3}}
\put(62,5){\line(0,1){13.5}}
\put(68,3.5){\line(0,1){18}}
\put(68,24.5){\line(0,1){3}}
\put(68,30.5){\line(0,1){3}}
\put(68,36.5){\line(0,1){3}}
\put(76,3.5){\line(0,1){18}}
\put(76,24.5){\line(0,1){3}}
\put(76,30.5){\line(0,1){10.5}}
\put(82,5){\line(0,1){19.5}}
\put(88,3.5){\line(0,1){12}}
\put(88,18.5){\line(0,1){9}}
\put(88,30.5){\line(0,1){10.5}}

\put(10,35){\line(1,0){4.5}}
\put(17.5,35){\line(1,0){9}}
\put(29.5,35){\line(1,0){5}}
\put(37.5,35){\line(1,0){9}}
\put(49.5,35){\line(1,0){5}}
\put(57.5,35){\line(1,0){9}}
\put(69.5,35){\line(1,0){23}}
\put(16,35){\circle{3}}
\put(28,35){\circle{3}}
\put(36,35){\circle{3}}
\put(48,35){\circle{3}}
\put(56,35){\circle{3}}
\put(68,35){\circle{3}}
\put(76,35){\circle*{2.5}}
\put(88,35){\circle*{2.5}}

\put(10,29){\line(1,0){4.5}}
\put(17.5,29){\line(1,0){17}}
\put(37.5,29){\line(1,0){17}}
\put(57.5,29){\line(1,0){9}}
\put(69.5,29){\line(1,0){5}}
\put(77.5,29){\line(1,0){9}}
\put(89.5,29){\line(1,0){3}}
\put(16,29){\circle{3}}
\put(28,29){\circle*{2.5}}
\put(36,29){\circle{3}}
\put(48,29){\circle*{2.5}}
\put(56,29){\circle{3}}
\put(68,29){\circle{3}}
\put(76,29){\circle{3}}
\put(88,29){\circle{3}}

\put(10,23){\line(1,0){4.5}}
\put(17.5,23){\line(1,0){9}}
\put(29.5,23){\line(1,0){17}}
\put(49.5,23){\line(1,0){5}}
\put(57.5,23){\line(1,0){9}}
\put(69.5,23){\line(1,0){5}}
\put(77.5,23){\line(1,0){15}}
\put(16,23){\circle{3}}
\put(28,23){\circle{3}}
\put(36,23){\circle*{2.5}}
\put(48,23){\circle{3}}
\put(56,23){\circle{3}}
\put(68,23){\circle{3}}
\put(76,23){\circle{3}}
\put(88,23){\circle*{2.5}}

\put(10,17){\line(1,0){16.5}}
\put(29.5,17){\line(1,0){5}}
\put(37.5,17){\line(1,0){17}}
\put(57.5,17){\line(1,0){29}}
\put(89.5,17){\line(1,0){3}}
\put(16,17){\circle*{2.5}}
\put(28,17){\circle{3}}
\put(36,17){\circle{3}}
\put(48,17){\circle*{2.5}}
\put(56,17){\circle{3}}
\put(68,17){\circle*{2.5}}
\put(76,17){\circle*{2.5}}
\put(88,17){\circle{3}}

\put(10,11){\line(1,0){24.5}}
\put(37.5,11){\line(1,0){9}}
\put(49.5,11){\line(1,0){5}}
\put(57.5,11){\line(1,0){35}}
\put(16,11){\circle*{2.5}}
\put(28,11){\circle*{2.5}}
\put(36,11){\circle{3}}
\put(48,11){\circle{3}}
\put(56,11){\circle{3}}
\put(68,11){\circle*{2.5}}
\put(76,11){\circle*{2.5}}
\put(88,11){\circle*{2.5}}

\put(10,5){\line(1,0){82.5}}

\put(94,41){\line(1,0){2}}
\put(94,35){\line(1,0){2}}
\put(94,29){\line(1,0){2}}
\put(94,23){\line(1,0){2}}
\put(94,17){\line(1,0){2}}
\put(94,11){\line(1,0){2}}
\put(94,5){\line(1,0){2}}
\put(97.5,41){\line(1,0){2}}
\put(97.5,35){\line(1,0){2}}
\put(97.5,29){\line(1,0){2}}
\put(97.5,23){\line(1,0){2}}
\put(97.5,17){\line(1,0){2}}
\put(97.5,11){\line(1,0){2}}
\put(97.5,5){\line(1,0){2}}
\put(101,41){\line(1,0){2}}
\put(101,35){\line(1,0){2}}
\put(101,29){\line(1,0){2}}
\put(101,23){\line(1,0){2}}
\put(101,17){\line(1,0){2}}
\put(101,11){\line(1,0){2}}
\put(101,5){\line(1,0){2}}
\put(104.5,41){\line(1,0){4.5}}
\put(104.5,35){\line(1,0){4.5}}
\put(104.5,29){\line(1,0){4.5}}
\put(104.5,23){\line(1,0){4.5}}
\put(104.5,17){\line(1,0){4.5}}
\put(104.5,11){\line(1,0){4.5}}
\put(104.5,5){\line(1,0){4.5}}
\put(10,11){\oval(3,3)[bl]}
\put(8.5,11){\line(0,1){7.5}}
\put(8.5,21.5){\line(0,1){7.5}}
\put(10,29){\oval(3,3)[tl]}
\put(7,18.5){\oval(3,3)[tr]}
\put(7,21.5){\oval(3,3)[br]}

\put(109,11){\oval(3,3)[br]}
\put(110.5,11){\line(0,1){13.5}}
\put(110.5,27.5){\line(0,1){13.5}}
\put(109,41){\oval(3,3)[tr]}
\put(112,24.5){\oval(3,3)[tl]}
\put(112,27.5){\oval(3,3)[bl]}
\end{picture}
\caption{The cloning stage for the $2\rightarrow 4$ quantum cloning
circuit. Only four permutations $\ket{000011}\rightarrow \ket{001001}$,
$\ket{000100}\rightarrow \ket{001010}$, $\ket{000000}\rightarrow\ket{000011}$,
and $\ket{110011}\rightarrow \ket{110101}$ are shown. Each of these
permutations consists of three multi-qubit control operations.
This figure represents only one possible example because the
order of permutation is not unique.}\label{2to4permnet}
\end{figure}

One possible permutation of bases that achieves the transformation
(\ref{2to4uqcm}) from $\ket{\psi}^{\mathrm{prep}}$ in Eq. (\ref{2to4preptrans})
is shown in Table \ref{2to4perm}. This can be implemented by the
quantum circuit in Figure \ref{2to4permnet}. With the quantum circuits
of Figures \ref{2to4prepnet} and \ref{2to4permnet}, the quantum cloning
transformation (\ref{2to4uqcm}) is carried out faithfully and the
upper three qubits in Figure \ref{2to4permnet} will be the clones, while
the next two qubits are the state of the cloner and the lowest
disentangled qubit is an ancilla which can be discarded.

\begin{table}
\caption{The permutations of bases to complete the transformation
Eq. (\ref{2to4uqcm}) when the circuit of the preparation is
given by Figure \ref{2to4prepnet}. Many of them are omitted in this table
since they are almost obvious by comparing Eq. (\ref{2to4uqcm}) and Eq.
(\ref{2to4preptrans}).}\label{2to4perm}
\begin{center}
\begin{tabular}{ccc|ccc}\hline
 Initial basis & $\rightarrow$ & Final basis & Initial basis & $\rightarrow$ & Final basis\\ \hline 
$\ket{000000}$ & $\rightarrow$ & $\ket{000011}$ & $\ket{110000}$ & $\rightarrow$ & $\ket{111100}$\\
$\ket{000001}$ & $\rightarrow$ & $\ket{000101}$ & $\ket{110001}$ & $\rightarrow$ & $\ket{111001}$\\
$\ket{000010}$ & $\rightarrow$ & $\ket{000110}$ & $\ket{110010}$ & $\rightarrow$ & $\ket{111010}$\\
$\ket{000011}$ & $\rightarrow$ & $\ket{001001}$ & $\ket{110011}$ & $\rightarrow$ & $\ket{110101}$\\
$\vdots$ & $\vdots$ & $\vdots$ & $\vdots$ & $\vdots$ & $\vdots$\\
$\ket{001101}$ & $\rightarrow$ & $\ket{101000}$ & $\ket{111101}$ & $\rightarrow$ & $\ket{101011}$\\
$\ket{001110}$ & $\rightarrow$ & $\ket{110000}$ & $\ket{111110}$ & $\rightarrow$ & $\ket{110011}$\\
\hline
\end{tabular}
\end{center}
\end{table}

We can now estimate the number of CNOTs in the whole cloning circuit.
In \cite{barenco95}, the number of basic operations, i.e. single qubit
operations and CNOTs, to simulate a multi-qubit controlled operation
has been given as $O(n^2)$ for a controlled-$U$ gate with $n-1$ control
qubits and one auxiliary qubit. Therefore, the upper bound
for the number of CNOTs in the cloning stage is the order
of the number of bases involved, which is approximately twice of the
number of amplitudes produced in the preparation stage, multiplied by
the square of the number of qubits. As the number of the bases is given
by the {\small RHS} of Eq. (\ref{cond}), $O(2^{2M}M^{-\half})$, and there are
$O(2^{2(M-N)}(M-N)^2)$ CNOTs in the preparation stage, a quantum
cloning circuit of the type of Figure \ref{n2mnet} contains
$O(2^{2M}(M-N)^2(2^{-2N}+M^{-\half}))$ CNOTs at most in total.

This circuit looks rather inefficient especially because of the
preparation stage. The inefficiency comes partly from our requirement
that the preparation stage should be able to generate arbitrary superpositions.
If both $N$ and $M$ can be fixed and we do not have to control the parameters
of each single qubit gate, the task of the preparation stage is
much easier and the configuration of its circuit may well be much simpler
and more efficient.

The cloning stage is also inefficient due to the complexity of the network
for a general permutation. Despite the linear dependence on $M$ in the case
of $1\rightarrow M$ UQCM, the number of CNOTs grows
exponentially as $M$ increases when there are $N$ inputs, since each
permutation is done one by one in our circuit. It might be possible to
find a more efficient circuit, however, we have not succeeded.

\section{Decoherence Time and Cloning Time}\label{dec_time}
With the number of CNOT gates estimated in the previous
section, we can now compare the cloning time $T$ with the decoherence
time $\tau_{\mathrm{dec}}$. We focus on a quantum computing realization
which makes use of cooled trapped ions \cite{cirac95}. In the following,
we assume that spontaneous emissions are the only source of decoherence,
and we only discuss the process without error correction codes. Going
beyond these constraints will be discussed elsewhere.

The Hamiltonian operator for a two-level ion of mass $m$, interacting
with a phonon as a result of the centre-of-mass (c.m.) motion with
frequency $\nu$, is then given by
\cite{cirac95, martin9697},
\begin{equation} \label{ham01}
 H=\frac{\eta}{\sqrt{2M-N}}\frac{\Omega_1}{2}[\ket{1}\bra{0}a+\ket{0}\bra{1}
a^{\dag}],
\end{equation}
where $\eta=(2\pi /\lambda)\sqrt{\hbar/2m\nu}$ is the Lamb-Dicke parameter,
$\Omega_1$ is the Rabi frequency of the $0\leftrightarrow 1$ transition
with 0 and 1 denoting the ground and the excited states of the ion.
The $a$ and $a^{\dag}$ are the annihilation and creation operators
of the phonon. As in the original proposal \cite{cirac95}, we assume that
qubits are encoded in the internal state of ions and the phonons are used
as the information bus. The denominator $\sqrt{2M-N}$ is a consequence of
the fact that an $N\rightarrow M$ UQCM network needs $2M-N$ qubits (ions). 

The elementary time step for a CNOT gate with this system
can be written as
\begin{equation} \label{tauel}
 \tau_{\mathrm{el}}\simeq\frac{4\pi\sqrt{2M-N}}{\eta\Omega_1}.
\end{equation}
The total processing time for cloning is
\begin{equation} \label{tclo01}
 T\simeq \frac{4\pi\sqrt{2M-N}}{\eta\Omega_1}\epsilon 2^{2M+2}(M-N)^2
 \left(\frac{1}{2^{2N}}+\frac{1}{\sqrt{\pi M}}\right),
\end{equation}
where $\epsilon$ is a some proportion factor. Here, we use the number of
CNOTs in the preparation stage, because it is usually dominant
over that of CNOTs in the cloning stage, as mentioned above. Also, we
assume that all the CNOTs are performed sequentially one by one, despite
the possibility of building up a circuit which performs several operations
in parallel. Thus, the discussion below gives only a naive upper bound for
the number of cloneable qubits.
To minimize $T$, we wish to increase
the value for $\Omega_1$, which is related to the decay rate of the
excited state by \cite{martin9697, bez00}
\begin{equation} \label{rabianddecay}
 \frac{\Omega_1^2}{\Gamma_1}=\frac{6\pi c^3 \epsilon_0}{\hbar\omega_1^3}E^2,
\end{equation}
where $E, c, \epsilon_0$, and $\omega_1$ are the electric field strength
of the laser, the speed of light, the permittivity of vacuum, and the
transition frequency between the states 0 and 1, respectively.
Clearly, we would like to minimize the decoherence effects of $\Gamma_1$
by going to a metastable level, but then $\Omega_1$ is also small.
We can increase the Rabi frequency $\Omega_1$ by increasing the
electric field strength of the laser $E$, but it would then cause transitions
to other higher levels which may be more rapidly decaying than state 1,
and eventually spontaneous emissions which
destroy the coherence of the system. Let $\ket{2}$ represent all other
auxiliary levels which are coupled to the ground state $\ket{0}$.

\begin{figure} 
 \begin{center}
  \includegraphics[scale=0.6]{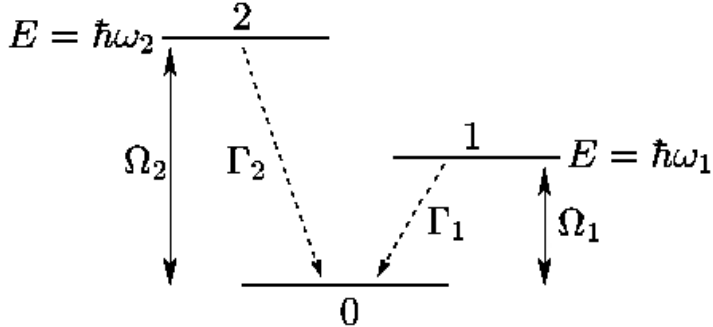}
  \caption{Three-level model of the ions used in a quantum cloner.
The transition between 0 and 1 represents the qubit. An external
laser drives this transition with Rabi frequency $\Omega_1$. The laser
inevitably couples level 0 to other non-resonant levels such as level 2.
The Rabi frequency of this transition is $\Omega_2$ and level 2 decays
with a rate $\Gamma_2$.}\label{threelevels} 
 \end{center}
\end{figure}

As a stronger laser increases the population in the auxiliary level,
it increases the rate of spontaneous emissions from level 2. Thus,
we compute the probability of an emission from either level 1 and 2,
and then minimize this probability to have an intensity-independent
limit to the number of output qubits under the effect of
spontaneous emission \cite{bez00}.

The probability of a spontaneous emission from the upper level of the
qubit during the cloning process is
\begin{equation} \label{prob1}
 p_{1\rightarrow 0}=\half 2\Gamma_1 (2M-N)T.
\end{equation}
The factor $\half$ is present because we can assume that on average
half of the qubits are in the upper level during the whole cloning
period $T$. Because the auxiliary level is populated only when interacting
with the laser, the probability of a spontaneous emission from level 2
is written as
\begin{equation} \label{prob2}
 p_{2\rightarrow 0}=\frac{\Omega_2^2}{8\Delta_2^2}2\Gamma_2 T,
\end{equation}
where $\Delta_2$ is the detuning between the frequency of the laser
and that of the transition $0\leftrightarrow 1$. We now obtain the total
probability of spontaneous emission
\begin{eqnarray} \label{ptot}
 p_{\mathrm{total}} &=& p_{1\rightarrow 0}+p_{2\rightarrow 0} \nonumber\\
 &=& \frac{4\pi\epsilon\sqrt{(2M-N)^3 \Gamma_1}}{\eta}2^{2M+2}(M-N)^2
 \left(\frac{1}{2^{2N}}+\frac{1}{\sqrt{\pi M}}\right) \nonumber\\
 & & \times\left[\frac{1}{x}+\frac{1}{2M-N}\left(\frac{\omega_1}{\omega_2}\right)^3
 \frac{\Gamma_2^2}{4\Delta_2^2\Gamma_1}x\right],
\end{eqnarray}
where $x=\Omega_1/\sqrt{\Gamma_1}$ and we used
\begin{equation}
 \frac{\Omega_1^2}{\Gamma_1}=\left(\frac{\omega_2}{\omega_1}\right)^3
 \frac{\Omega_2^2}{\Gamma_2}, \nonumber
\end{equation}
which is derived from Eq. (\ref{rabianddecay}). The minimum value
of $p_{\mathrm{total}}$ with respect to $x$ is
\begin{equation} \label{pmin}
 p_{\mathrm{min}}=\frac{\pi\epsilon}{\eta}
 \left(\frac{\omega_1}{\omega_2}\right)^{\frac{3}{2}}\frac{\Gamma_2}{\Delta_2}
 2^{2M+4}(2M-N)(M-N)^2\left(\frac{1}{2^{2N}}+\frac{1}{\sqrt{\pi M}}\right).
\end{equation}
In order for the cloning process to be successful, $p_{\mathrm{min}}$
should be very small compared with unity. This requirement gives the
upper bound on $M$ for a given $N$ in the expression of
\begin{eqnarray} \label{mmax1}
 2^{2M_{\mathrm{max}}}(2M_{\mathrm{max}}-N)(M_{\mathrm{max}}-N)^2
 \left(\frac{1}{2^{2N}}+\frac{1}{\sqrt{\pi M_{\mathrm{max}}}}\right) \nonumber\\
 \simeq \frac{\eta}{2^4\pi\epsilon}
 \left(\frac{\omega_2}{\omega_1}\right)^{\frac{3}{2}}\frac{\Delta_2}{\Gamma_2}.
\end{eqnarray}

Table \ref{tabions} shows the results of the actual calculations of the
{\small RHS} of Eq. (\ref{mmax1}) for those ions often utilized in trap
experiments along with their numerical data. Assuming $\epsilon=100$ \cite{rough},
$\eta=0.01$ and $\Delta_2=10^{13}\:\, [s^{-1}]$, we find that
Eq. (\ref{mmax1}) has no solutions for $M\ge 2$ as the {\small LHS} of
Eq. (\ref{mmax1}) is equal to 31.2 when $N=1$ and $M=2$.
However, if we consider only the $1\rightarrow 2$
cloning network, where the number of CNOTs is six according to
Figure \ref{uqcmnet}(a), the probability of spontaneous emission during
the whole process becomes rather small. For example, $p_{\mathrm{min}}=0.062$
for $\mathrm{Ca^+}$ and $p_{\mathrm{min}}=0.017$ for $\mathrm{Ba^+}$, therefore
the cloning may be successful.

\begin{table}
\caption{Atomic data of several possible systems for a quantum cloning
machine and the {\small RHS} of Eq. (\ref{mmax1}), which should be smaller
than the {\small LHS}. We can see that Eq. (\ref{mmax1}) has no solutions
for $N>1$ and $M>2$, as its {\small LHS}, which is equal to 31.15 when $N=1$
and $M=2$, is always larger than the {\small RHS} shown in this table.
Atomic data are taken from [21].}\label{tabions}
\begin{center}
\begin{tabular}{c||c|c|c}\hline
 Ion & $\mathrm{Ca^+}$ & $\mathrm{Hg^+}$ & $\mathrm{Ba^+}$ \\ \hline 
 level 0 & $4s^{\,2}\!S_{1/2}$ & $5d^{\,10}6s^{2\:\,2}\!S_{1/2}$ & $6s^{\,2}\!S_{1/2}$ \\
 level 1 & $3d^{\,2}\!D_{5/2}$ & $5d^{\,9}6s^{2\:\,2}\!D_{5/2}$ & $5d^{\,2}\!D_{5/2}$ \\
 level 2 & $4s^{\,2}\!P_{3/2}$ & $5d^{\,10}6p^{2\:\,2}\!P_{1/2}$ & $6s^{\,2}\!P_{3/2}$ \\
 \hline
 $\omega_1 \:[s^{-1}]$ & $2.62\times 10^{15}$ & $6.7\times 10^{15}$ &
 $1.07\times 10^{15}$ \\
 $\omega_2 \:[s^{-1}]$ & $4.76\times 10^{15}$ & $11.4\times 10^{15}$ &
 $4.14\times 10^{15}$ \\
 $\Gamma_2 \:[s^{-1}]$ & $67.5\times 10^6$ & $5.26\times 10^8$ &
 $58.8\times 10^6$ \\ \hline
 {\small RHS} of Eq. (\ref{mmax1})\:$(\eta=0.01)$ & $0.72$ & $0.084$ & $2.58$ \\
 \hline 
\end{tabular}
\end{center}
\end{table}

Overall, we see that even for a small number of outputs copied from
one input the decoherence due to spontaneous emissions plays a critical role.
If we make an optimistic assumption for $\eta$, as $\eta=1.0$,
$1\rightarrow 2$ cloning with $\mathrm{Ca^+}$, $1\rightarrow 2$ and
$2\rightarrow 3$ with $\mathrm{Ba^+}$ would become possible.

If we are allowed to have one more auxiliary qubit, the number of
basic operations to simulate a $(n-2)$-controlled operation can be
reduced from $O(n^2)$ to $O(n)$ \cite{barenco95}. Replacing the
corresponding factors in the above calculations shows that, with
$\eta=1.0$, $1\rightarrow 2$, $1\rightarrow 3$, and $2\rightarrow 3$
cloning may be possible with $\mathrm{Ba^+}$ as well as $1\rightarrow 2$
and $1\rightarrow 3$ cloning with $\mathrm{Ca^+}$.

\section{Summary}
We have first investigated a possible method to construct an $N\rightarrow M$
quantum cloning circuit in order to estimate the number of CNOT
gates in it. 
We have estimated the number of CNOTs as $O(2^{2M}(M-N)^2(2^{-2N}+M^{-\half}))$,
with $N$ and $M$ the numbers of input and output qubits, respectively.
Therefore, the number of gates in the quantum cloning circuit of the type of
Figure \ref{uqcmnet}(b) or Figure \ref{n2mnet} is always exponential
with respect to the number of output qubits. 

With the circuit complexity we obtained, it has been shown that the
quantum cloning may be implemented by using the system of trapped ions
for only a few combinations of small $N$ and $M$, when the system is not
immune to decoherence due to spontaneous emissions, provided sophisticated
quantum error correction codes are not used. As spontaneous emissions are
the only source of decoherence we have considered here, the number of possible
combinations of $N$ and $M$ might be lowered further by taking into account
of other effects, such as phonon decoherence \cite{garg96}, the random phase
fluctuations of the lasers, the heating of the ions' vibrational motion
\cite{hughes96, murao98, james98, schneider98}.

Nevertheless, unlike the case of factorization by Shor's algorithm
\cite{martin9697}, producing many clones is not necessarily what we
expect from the quantum cloning. Since even a few copied
qubits may be useful in quantum information processing \cite{galvao00}, our
results should not be interpreted too pessimistically. Furthermore,
the use of quantum error correction codes will surely ease the
condition for upper bounds.

In \cite{hillery02}, it was shown that a quantum information distributor,
which is a modification of the quantum cloning circuit, can be used
as a universal programmable quantum processor in a probabilistic regime.
From this point of view, our estimation on the upper bound for the clones
implies physical bounds on the realization of such a processor
with trapped ions.

One interesting subject we have not considered here is the effect
of decoherence on the quality of the clones. If we allow a processing
time which is longer than the decoherence time, the fidelity between the
input and output states will surely be lower than what we expect from
Eq. (\ref{guqcm}). We might be able to find a tradeoff between the
fidelity and the number of cloneable qubits to have ``useful''
clones in terms of practical quantum information processing.

\section*{Acknowledgement}
We are very grateful to V. \Buzek for a careful reading of the manuscript and
helpful suggestions. This work was supported in part by the European
Union IST Network QUBITS. K.M. acknowledges financial support by Fuji
Xerox Co., Ltd. in Japan.


\begin{thebibliography}{99}
\bibitem{rough}
This is a rough estimation based on Corollary 7.4 in \cite{barenco95}.
\end{thebibliography}
\end{document}